\documentclass[aps,prl,twocolumn,groupedaddress,superscriptaddress]{revtex4-2}
\usepackage{graphicx}
\usepackage{amsmath}
\usepackage{epstopdf}
\usepackage{float}
\usepackage{color}
\usepackage{appendix}

\usepackage{multirow}
\usepackage{booktabs}
\usepackage{longtable}

\DeclareMathOperator{\arctanh}{arctanh}

\newcommand{\meanvalue}[1]{\langle#1\rangle}
\newcommand{\ket}[1]{|#1\rangle}
\newcommand{\bra}[1]{\langle#1|}

\newcommand{\Jz}{\hat{J}_z}
\newcommand{\Jx}{\hat{J}_x}
\newcommand{\Jy}{\hat{J}_y}
\newcommand{\Jplus}{\hat{J}_+}
\newcommand{\Jminus}{\hat{J}_-}

\newcommand{\Hop}{\hat{H}}
\newcommand{\Sop}{\hat{S}}

\newcommand{\aop}{\hat{a}}
\newcommand{\adag}{\hat{a}^{\dagger}}
\newcommand{\adagsq}{\hat{a}^{\dagger 2}}

\newcommand{\bop}{\hat{b}}
\newcommand{\bdag}{\hat{b}^{\dagger}}

\newcommand{\dop}{\hat{d}}
\newcommand{\opddag}{\hat{d}^{\dagger}}

\newcommand{\betasqr}{\hspace{5pt} \sqrt[]{N-\beta^{2}} \hspace{5pt}}

\newcommand{\gprime}{g_{\beta}}

\newcommand{\Ko}{\hat{K}_0}
\newcommand{\Kp}{\hat{K}_+}
\newcommand{\Km}{\hat{K}_-}

\newcommand{\Kop}{\hat{K}_0\sp{\prime}}
\newcommand{\Kpp}{\hat{K}_+\sp{\prime}}
\newcommand{\Kmp}{\hat{K}_-\sp{\prime}}

\begin{document}

\title{Stark-induced tunable phase transition in the two-photon Dicke-Stark model}

\author{Cui-Lu Zhai}
\affiliation{Institute for Quantum Science and Technology, College of Science, NUDT, Changsha 410073, China}
\affiliation{Hunan Key Laboratory of Mechanism and technology of Quantum Information, Changsha, 410073, China}
\affiliation{Department of Maths and Physics, Hunan Institute of Engineering, Xiangtan 411104, China}
\author{Wei Wu}
\affiliation{Institute for Quantum Science and Technology, College of Science, NUDT, Changsha 410073, China}
\affiliation{Hunan Key Laboratory of Mechanism and technology of Quantum Information, Changsha, 410073, China}
\author{Chun-Wang Wu}
\email{cwwu@nudt.edu.cn}
\affiliation{Institute for Quantum Science and Technology, College of Science, NUDT, Changsha 410073, China}
\affiliation{Hunan Key Laboratory of Mechanism and technology of Quantum Information, Changsha, 410073, China}
\author{Ping-Xing Chen}
\email{pxchen@nudt.edu.cn}
\affiliation{Institute for Quantum Science and Technology, College of Science, NUDT, Changsha 410073, China}
\affiliation{Hunan Key Laboratory of Mechanism and technology of Quantum Information, Changsha, 410073, China}

\date{\today}

\begin{abstract}

We theoretically investigate the superradiant phase transition (SPT) in the two-photon Dicke-Stark model, which incorporates both Rabi and Stark coupling.
By introducing a Stark coupling term, we significantly reduce the critical Rabi coupling strength required to achieve the SPT, enabling it to occur even in strong coupling regimes. Using mean-field theory, we derive the conditions for the SPT and show that it exhibits a second-order phase transition. Surprisingly, we demonstrate that the transition point can be widely tuned by the Stark coupling strength. The signatures of these Stark-tunable SPT points are manifested through atomic averages. When quantum fluctuations are included, the spin-squeezing distributions also reveal the effects of Stark-tunable SPT points. In addition, we propose an experimentally feasible realization using an ion trap system driven by three lasers. Our scheme enables optical switching between normal and superradiant phases through pump field intensity modulation, where the Stark coupling coefficient serves as the optically tunable parameter. Our results offer a new approach to engineer the SPT, extending superradiance-based quantum technologies beyond the ultrastrong coupling regime.
\end{abstract}


\maketitle

\noindent{\it Introduction. }
Quantum phase transitions~\cite{Sachdev:2017,Xu:2011,Leonard:2017} mark abrupt changes between distinct quantum phases at zero temperature, governed by quantum fluctuations. A paradigmatic example is the superradiant phase transition (SPT)~\cite{Hepp:1973, Hioe:1973, Wang:1973} in the Dicke model, where a second-order phase transition occurs when the qubit-cavity coupling surpasses a critical threshold~\cite{Emary:2003}. Below this threshold, the system remains in a symmetric normal phase (NP) with the cavity field in the vacuum state and atoms in their ground states; above it, symmetry breaking leads to a macroscopic photon population and collective atomic excitation---the superradiant phase (SP).
Experimental advances have demonstrated nonequilibrium SPT in diverse quantum platforms, including Bose-Einstein condensates~\cite{Baumann:2010,Klaers:2010,Baumann:2011,Klinder:2015} and trapped ion systems~\cite{Safavi-Naini:2018}. These observations have stimulated theoretical investigations into how external driving fields and dissipation govern quantum phase~\cite{Diehl:2010,Bastidas:2012,Kirton:2017,Soriente:2018,Gutierrez-Jauregui:2018,Soriente:2020,Kirton:2018}. Many variants of the SPT model have also been theoretically developed~\cite{Nagy:2010,Baksic:2014,Liu:2011,Liu:2013,Lu:2018,Felicetti:2020,Samimi:2022}. Recent work has extended the SPT to few-body systems including the Rabi~\cite{Liu:2017,Puebla:2017,Zhang:2021,Bakemeier:2012,Hwang:2015,Hwang:2018,Chen:2020,Stransky:2021} model, Tavis-Cummings~\cite{Hwang:2016,Carmichael:2015,Fink:2017} model, a single qubit interacting with a single oscillator~\cite{Ashhab:2013} and even a nonlinear Kerr resonator~\cite{Bartolo:2016}. Very recently, the quantum phase transition of the Rabi model has been simulated with trapped ions~\cite{Lv:2018,Cai:2021}, and that of the Tavis-Cummings model~\cite{Feng:2015} implemented using superconducting qubits.

The Dicke SPT emerges in the thermodynamic limit through the cooperative interplay between quantum fluctuations and collective interactions,  and provides the foundational framework for understanding quantum phase transitions. Note that the original work on Dicke phase transition requires ultrastrong coupling regimes, where the collective coupling strength approaches the cavity frequency. While recent advances in superconducting circuits and ion traps have enabled the achievement of such ultrastrong coupling~\cite{Niemczyk:2010,Langford:2017,Braumuller:2017,Forndiaz:2010,Peterson:2019,Kockum:2019}, the critical parameter regime required for equilibrium SPT remains challenging to satisfy with current cavity QED technologies. This limitation has driven the search for alternative models and mechanisms that can achieve superradiance under more feasible conditions~\cite{Zou:2013,Mazza:2019,Zhang:2020,Huang:2023,Sadhasivam:2024,Xie:2025}.

In this work, we present a novel approach to study SPT in the two-photon Dicke-Stark model. By introducing the Stark coupling term, we demonstrate that this model exhibits a second-order phase transition without ultrastrong coupling. Specifically, increasing the Stark coupling strength reduces the critical coupling required for the transition, allowing it to occur even in the strong coupling regime. This breakthrough extends superradiance-based quantum metrology~\cite{Bina:2016,Ilias:2022,Zhang:2024,Montenegro:2024,Ilias:2024,Yousefjani:2025} beyond the ultrastrong coupling limit, opening new possibilities for quantum sensing across a wider range of coupling strengths.
Within mean-field theory, where atomic fluctuations are neglected, we demonstrate that the SPT critical point can be continuously tuned through Stark coupling, as evidenced by atomic averages. Beyond this approximation, our quantum fluctuation analysis can also reveal these Stark-tuned SPT critical points through their characteristic atomic spin-squeezing distributions with different profiles.
Notably, we also accounts for and avoids the spectral collapse condition typically associated with the two-photon Dicke model.
\begin{figure*}[htbp]
	\begin{center}
		\includegraphics[width=1.8\columnwidth, angle=0]{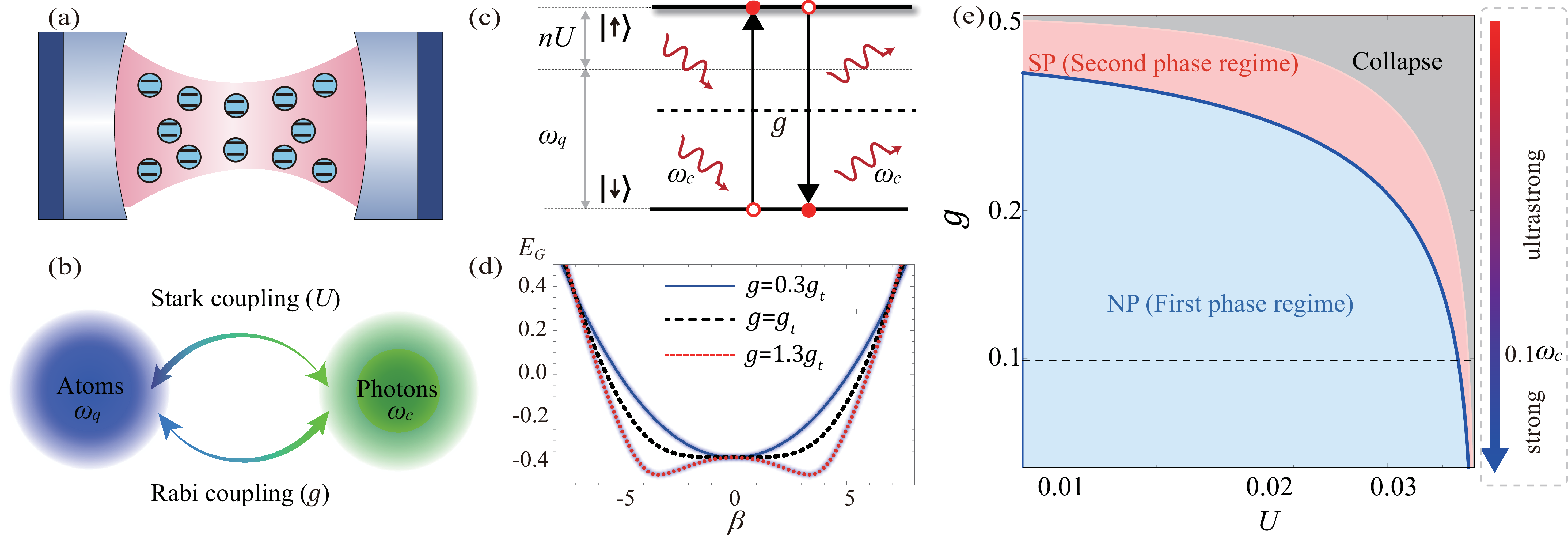}
		\caption{The model. (a) Schematic of the two-photon Dicke model comprising an ensemble of $N$ qubits coupled to a bosonic mode. (b) Sketch for qubits-resonator coupling diagram. (c) Two-photon creation and annihilation processes. Note that due to the Stark term, the atomic energy levels shift. (d) The ground-state energy in the two-photon Dicke-Stark model as a function of the order parameter $\beta$ defined in Eq.~(\ref{beta}) for different Rabi coupling strength. The blue solid line and red dotted line correspond to the NP and SP, respectively. The black dashed line represents the phase boundary. Here, $U=0.0168\omega_{c}$. (e) Phase diagram. Increasing the Stark coupling strength $U$ suppresses the critical Rabi coupling strength, enabling access to the SPT from ultrastrong to strong coupling regimes. This tunability broadens the experimental parameter space for quantum metrology applications. Note that the gray area corresponds to the spectral collapse region to be avoided. Other parameters: $\omega_{q}=0.015\omega_{c}$ and $N=50$.}\label{The model}
	\end{center}
\end{figure*}

Additionally, we propose an experimentally feasible implementation using a trapped-ion system driven by three lasers. In our scheme, the optical control of the Stark coupling is achieved through laser intensity modulation, enabling an optical switching between the NP and SP via pump field adjustment. In addition, all system parameters are experimentally accessible and independently tunable. Unlike methods that rely on Rabi coupling adjustments, this trapped-ion platform offers more feasible and flexible optical control, making it an ideal system for studying the SPT and advancing precision quantum metrology.

\noindent{\it The model. }
Let us consider an ensemble of $N$ qubits interacting with a bosonic mode through the addition of a Stark shift term to the two-photon Dicke model, as illustrated in Fig.~\ref{The model}(a). The system Hamiltonian is given by:
\begin{equation}
\Hop=\omega_{c} \adag \aop + \omega_q \Jz + \frac{g}{N}(\Jplus + \Jminus)(\aop^2 + \adagsq)+ U    \adag a\Jz,
\label{model DSM J}
\end{equation}
where $\hat{a}^\dagger $ and $\aop$ are the creation and annihilation operators of the bosonic mode, respectively. $\hat{J}_{\pm}=\sum_{j}{\hat{\sigma}_\pm}^{j}$ are the collective spin raising and lowering operators and $\hat{J}_{\alpha}=\frac{1}{2}\sum_{j}\hat{\sigma}^{j}_{\alpha}$ ($\alpha=x,y,z$) are the macroscopic spin operators. $\omega_q$ is the transition frequency of the qubits. $\omega_{c}$ is the frequency of the bosonic field. The $U$ term represents the Stark shift for nonlinear atom-cavity interactions. $g/N$ denotes the individual Rabi coupling strength. The qubits-resonator coupling of the system is illustrated in Fig.~\ref{The model}(b). The photon creation and annihilation processes involved in the two-photon Dicke-Stark model are explicitly shown in the Fig.~\ref{The model}(c). Here, $\aop^2$ ($\adagsq$) annihilates (creates) a pair of photons in the cavity. Due to the Stark term, the energy levels of the atoms are modified. The Hamiltonian commutes with a generalized $Z_{4}$ parity operator $\hat{\Pi} $, defined as $\hat{\Pi}  =(-1)^{N}\otimes _{j=1}^{N}\hat{\sigma}_{z}^{(j)}e^{i\pi \adag \aop/2}$. The operator $\hat{\Pi}  $ has four eigenvalues: $\pm 1$ and $\pm i$. The $Z_{4}$ parity symmetry is expected to be spontaneously broken in the ground state during the super-radiant phase transition. For simplicity, we set $\omega_{c}=1$ in the following.

In the ultrastrong regime, the two-photon Dicke model shows spectral collapse~\cite{Travenec:2012,Felicetti:2015}---energy levels become continuous when the coupling strength reaches a critical value, specifically at $g=\omega_{c}/2$. Beyond this point, the ground state becomes no longer defined, making phase transitions meaningless. Typically, spectral collapse occurs beyond the phase transition point as the Rabi coupling strength increases. We study SPT in the two-photon Dicke-Stark model while ensuring to avoid spectral collapse.

\noindent{\it Phase diagram. } We investigate the phase diagram of the two-photon Dicke-Stark model using a mean-field approach~\cite{Baksic:2014}. The point here is to determine how the properties of the ground state evolve when $g$ increases. In the thermodynamic limit, we employ the Holstein-Primakoff transformation to express the collective spin operators in terms of bosonic operators: $\Jplus = \bdag\sqrt{N - \bdag\bop}$, $\Jminus = \sqrt{N - \bdag\bop}\bop$, and $\Jz = \bdag\bop - N/2$. Here, the bosonic operators satisfy the canonical commutation relation $[\bop, \bdag]=1$.
The bosonic modes are then displaced relative to their ground-state expectation values as: $\bop\rightarrow\beta + \dop$. The spin fluctuations obey with $[\dop, \opddag]=1$ and
\begin{equation}
\beta=\bra{GS}\bop \ket{GS},
\label{beta}
\end{equation}
where $\ket{GS}$ is the ground state. To zeroth order approximation, this yields: $\Hop = \omega_{\beta}\adag \aop + \gprime(\aop^2 + \adagsq) + \omega_q\beta^{2} - \omega_qN/2$, where $\omega_{\beta}=\omega_{c}+U(\beta^{2}-N/2)$, $\gprime=2g \beta \betasqr/ N$ and $\beta$  is taken to be real.
The Hamiltonian is quadratic in $\aop$ and can thus be diagonalized via Bogoliubov transformation to obtain the ground-state energy:
\begin{eqnarray}
E_G &=&\frac{\omega_{\beta}^2 - 4\gprime^2}{2\omega_{\beta}} \cosh(2r_{\beta})+\left( \omega_q -\frac{U}{2}\right)\beta^{2} \nonumber  \\
&&    - \frac{\omega_qN}{2} - \frac{\omega_{c}}{2} + \frac{UN}{4},
\end{eqnarray}
where the squeezing parameter is given by $r_{\beta}=\arctan(2g_{\beta}/\omega_{\beta})/2$. In Fig.~\ref{The model}(d), we plot the corresponding ground-state energy with respect to the order parameter $\beta$ for different Rabi coupling strengths. The transformation of the energy functional from a single to a double symmetric well suggests that the system is undergoing a quantum phase transition.
\label{SPT}

The ground state configuration is determined by energy minimization with respect to $\beta$, which serves as the order parameter. By minimizing $E_{G}$, we obtain the following results: Below the critical coupling ($g<g_t$), the energy minimum occurs at $\beta=0$, corresponding to a NP where the field remains in its vacuum state and the atoms are unexcited. Above the transition point($g>g_t$), the system enters a SP characterized by a two-fold degenerate ground state with $\beta=\pm|\beta|$, indicating spontaneous symmetry breaking. In this phase, the field evolves into a squeezed vacuum state whose squeezing direction depends crucially on the sign of $\beta$ while the system develops collective atomic excitations.
\begin{figure}[t]
    \centering
	\includegraphics[scale=0.6]{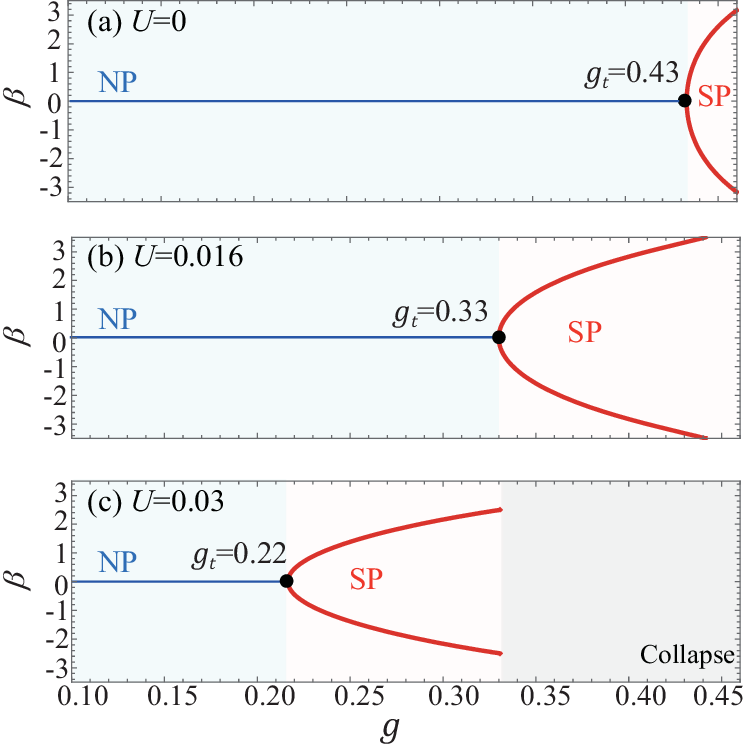}
    \caption{The order parameter $\beta$ in the two-photon Dicke-Stark model as a function of the Rabi coupling strength for different Stark coupling $U$. (a) $U=0$ and $g_{t}=0.43\omega_{c}$. (b) $U=0.0168\omega_{c}$ and $g_{t}=0.33\omega_{c}$. (c) $U=0.03\omega_{c}$ and $g_{t}=0.22\omega_{c}$. $\beta$ is zero in the NP and finite in the SP. Note that $\beta^{2}$ represents the mean value of atomic ground state. Other parameters are the same as in Fig.~\ref{The model}.}
\label{order parameter}
\end{figure}

In the case where the order parameter $\beta$ is real, the critical Rabi coupling strength for the phase transition can be derived as $g_{t} =\sqrt{ \omega_{c}\omega_{q}N - U \omega_{q} N^{2}/2}/2$ while the spectral collapse occurs at $g_{c} = \sqrt{\omega_{c}^{2}-U^{2} N^{2}/4}/2$. To properly investigate the phase transition in our systems, we must maintain the condition $g<g_{c}$ to avoid entering the collapsed regime. Figure~\ref{The model}(e) displays the phase diagram of the model in the mean-field approximation. Our analysis reveals that increasing the Stark coupling strength $U$ significantly reduces the critical Rabi coupling strength $g_t$, enabling the SPT to be accessed from the ultrastrong to strong coupling regimes. This tunability significantly expands the accessible parameter space for superradiance-based quantum metrology - a central result of our work.
\begin{figure}[t]
    \centering
	\includegraphics[scale=0.63]{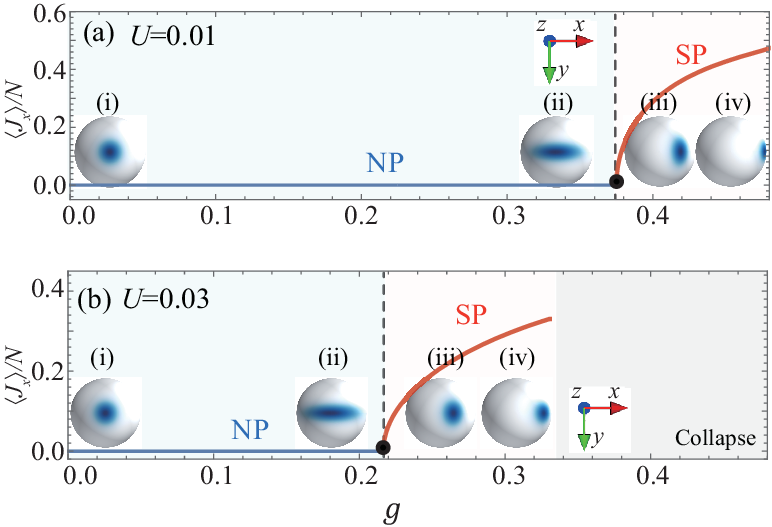}
    \caption{Mean value of $\Jx/N$ as a function of the Rabi coupling strength for (a) $U=0.01\omega_{c}$ and (b) $U=0.03\omega_{c}$, respectively. The inserted figures are schematic of representation spin-squeezing fluctuations of the collective angular momentum ($\hat{J}$) state in a generalized Bloch sphere. From left to right, four distinct regimes of the Rabi coupling strength were considered: (i) the weak-coupling regime ($g<g_{t}$), (ii) the transition-approaching regime ($g\lesssim g_{t}$), (iii) the superradiant regime ($g_{t}<g<g_{c}$) and (iv) the near-collapse regime ($g\lesssim g_{c}$). Other parameters are the same as in Fig.~\ref{The model}.}
\label{spin squeezing fluctuations}
\end{figure}

\noindent{\it Stark-induced tunable SPT point. }
To investigate the Stark-tunable SPT, we first employ mean-field theory to identify the critical phase boundary.
Figure~\ref{order parameter} shows the order parameter $\beta$ as a function of Rabi coupling strength for various Stark couplings ($U$).
Here,  $\beta^{2}$ represents atomic ground-state population, and the detailed formula of $\beta$ is provided in the Supplemental Material. For the $U=0$ case, a quantum phase transition occurs at $g_{t}=0.43\omega_{c}$, where $\beta$ bifurcates from zero (NP) to finite values (SP). Remarkably, introducing finite $U$ reduces the Rabi critical coupling: $g_{t}$ decreases to $0.33\omega_{c}$ at $U=0.0168\omega_{c}$ and further to $0.22\omega_{c}$ at $U=0.03\omega_{c}$. This demonstrates that Stark coupling provides an effective control mechanism for tuning the superradiance threshold.

However, mean-field theory neglects quantum fluctuations, which are essential for a complete characterization of the system. To go beyond this approximation, we now analyze spin-squeezing distributions which encode information about atomic fluctuations $\hat{d}$. Additionally, we introduce the SU(1,1) Lie algebra operators $\Ko=\frac{1}{2}(\adag \aop + \frac{1}{2}), \Kp=\frac{1}{2} \adagsq$, and $\Km=\frac{1}{2} \aop^2$, which satisfy the commutation relations: $ [\Ko,\hat{K}_{\pm}]=\pm \hat{K}_{\pm},$ and $\hspace{5pt} [\Kp, \Km]=-2\Ko$. Using Holstein-Primakoff and Schrieffer-Wolff transformations, we project the Hamiltonian into the lowest-energy eigenspace of $\Ko$. This procedure decouples the atomic and photonic degrees of freedom, yielding an effective Hamiltonian that describes exclusively the $d$ fluctuations above the ground state in both NP and SP. Up to a certain order of the small parameter of $1/\sqrt{N}$, the resulting formulation enables diagonalization of the effective Hamiltonian in each quantum phase (see Supplemental Material for details).
\begin{figure}[t]
    \centering
	\includegraphics[scale=0.3]{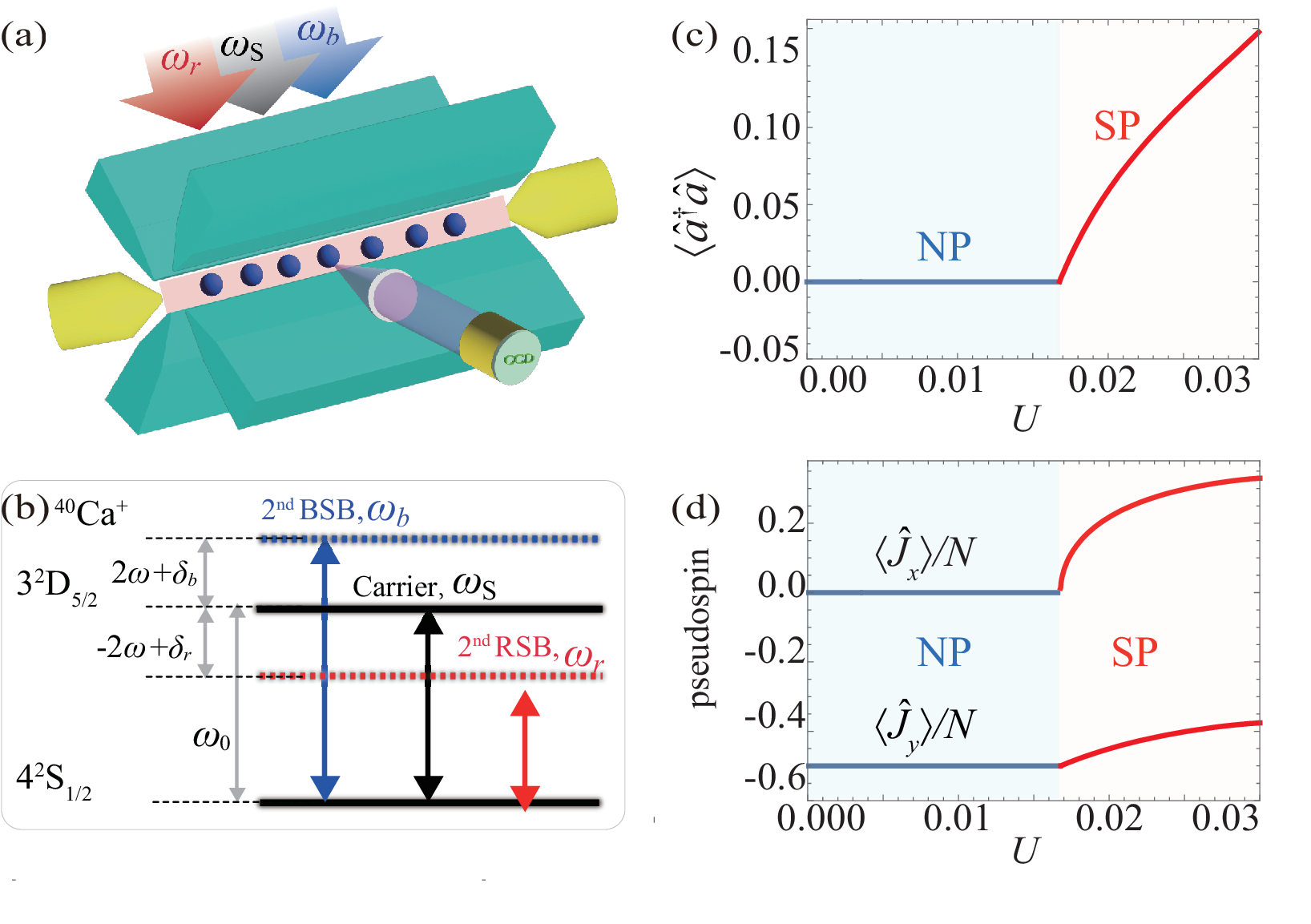}
    \caption{Implementation of our two-photon Dicke-Stark model in a trapped-ion setup. (a) $N$ trapped ions share a collective phonon mode, driven by three lasers: two second-sideband beams ($\omega_{r,b}$, detuned by $\delta_{r,b}$ ) and one carrier ($\omega_{\rm S}$), using the $4^{2}S_{1/2} \leftrightarrow 3^{2}D_{5/2}$  transition in $^{40} \text{Ca}^{+}$. (b) Corresponding energy level diagram. (c,d)
    Optical switching between the NP and SP is demonstrated through: (c) phonon number and (d) pseudospin as functions of Stark coupling strength. Other parameters are the same as in Fig.~\ref{The model}}
\label{ion trap}
\end{figure}

In the NP, we consider the case $\beta=0$: $\bop=\beta + \dop=\dop$. After decoupling the eigenspaces of $\Ko$ and projecting the system into the lowest-energy one, we derive the low-energy effective Hamiltonians as
\begin{equation}
\hat{H}_{NP}=\omega_q \opddag \dop - \frac{2 g^2 (\dop+\opddag)^2}{2\omega_{c}- N U}-\frac{\omega_q N}{2}.
\label{Hnp}
\end{equation}
The associated ground state for $\dop$ is a squeezed vacuum state, with squeezing parameter $r_{s}^{(1)}=\frac{1}{4}\ln[1-\frac{8g^{2}}{\omega_{q}N(2\omega_{c}-UN)}]$. Considering $N$ be infinite, this parameter is
negative, meaning that the squeezed quadrature is $\hat{P}_{d}$ instead of $\hat{X}_{d}$.

In the SP, $\beta\neq0$ and we express the field operator as $\bop=\beta + \dop$.  Following analogous procedures to the NP case, we derive the low-energy effective Hamiltonian for the fluctuation operator $\dop$:
\begin{equation}
\Hop_{SP} =\frac{\lambda_0}{4}  +\omega_{1}\opddag \dop +\omega_{2}(\dop+\opddag)^2,
\label{Hsp}
\end{equation}
which can be diagonalized via a Bogoliubov transformation with parameter $r_{s}^{(2)}=-\frac{1}{4}\ln(1+\frac{\omega_{2}}{\omega_{1}})$. The complete derivation and parameter definitions are provided in Supplemental Material.

Figure~\ref{spin squeezing fluctuations} displays the spin-squeezing fluctuations in the collective angular momentum $\hat{J}$ for different Stark coupling parameters $U$, where the signature of Stark-induced tunable SPT points can also be clearly observed through distinct spin-squeezing profiles. In the NP, the system exhibits $z$-axis polarization: $\meanvalue{\Jz}= -\frac{N}{2}$, and $\meanvalue{\Jx}=\meanvalue{\Jy}=0$; as $g$ increases, fluctuations of $\Jy$ are suppressed while fluctuations of $\Jx$ grow proportionally, diverging near the transition. In the SP, the pseudospin polarization evolves towards the $x$-axis: $\meanvalue{\Jz} = \lvert\beta\rvert^2-\frac{N}{2} $, $\meanvalue{\Jx} = \beta \sqrt{N-\beta^{2}} $ and $\meanvalue{\Jy} = 0$, meaning $\lvert\meanvalue{\Jz}\rvert$ decreases until it hits the value for $g=g_{c}$, and $\lvert\meanvalue{\Jx}\rvert$ increases at the same time. The squeezing effect becomes significantly diminished. Crucially, the SPT point $g_{t}$---separating these distinct fluctuation regimes---is Stark-tunable: stronger Stark coupling suppresses the critical Rabi coupling $g_{t}$, enabling experimental detection of the transition via spin-squeezing profiles. This demonstrates Stark coupling as a powerful tool for controlling quantum criticality and manipulating collective spin dynamics.

\noindent{\it Implementation with trapped ions. }
Various quantum optical platforms provide promising opportunities for implementing the model system demonstrating our scheme. Previous works have realized both one-photon~\cite{Pedernales:2015,Puebla:2016,Lv:2018} and two-photon~\cite{Felicetti:2015,Puebla:2017Protected} Rabi models in trapped ions. Specifically, Felicetti \textit{et al.}~\cite{Felicetti:2015} demonstrated a two-photon Dicke model using laser-driven ion chains, where the collective motion serves as the bosonic field. As a concrete example, we outline in the following a possible trapped-ion implementation of our two-photon Rabi-Stark model.

The interaction Hamiltonian for a single trapped ion with co-propagating laser beams (labeled $j$), in the rotating frame of $H_0=(\omega_0/2)\hat{\sigma}_z+\omega \hat{a}^\dagger \hat{a}$, is:
\begin{equation}\label{TIHamil}
\hat{H}^{I}=\sum_{j}\frac{\Omega_j}{2} \hat{\sigma}^+e^{i\eta[\hat{a}(t)+\hat{a}^\dag (t)]}e^{-i(\omega_j-\omega_0)t}e^{i\phi_j} +{\rm H.c.},
\end{equation}
where the Planck constant $\hbar=1$. We define $\hat{a}(t)=\hat{a} e^{-i\omega t}$ and $\hat{a}^\dag (t)=\hat{a}^\dag e^{i\omega t}$. $\hat{a}^\dagger $ and $\hat{a}$ are the creation and annihilation operators acting on vibrational phonons with frequency $\omega$. The two-level atom (level splitting $\omega_{0}$) is characterized by the Pauli matrices $\hat{\sigma}_{x,y,z}$ and $\hat{\sigma}_{\pm}=(\hat{\sigma}_{x}\pm i\hat{\sigma}_{y})/2$. The matrics satisfy the spin algebra $[\hat{\sigma}_{i} ,\hat{\sigma}_{j}]=2i \epsilon_{ijk}\hat{\sigma}_{k}$ with $i,j,k\in \{x,y,z\}$. The Rabi frequency $\Omega_j$ scales with the laser amplitude, while $\phi_j$ denotes its phase. The Lamb-Dicke parameter is expressed as $\eta = k_z \sqrt{\frac{\hbar}{2 m \omega}}$, where $k_z$ represents the component of the wavevector of the $j$-th laser field aligned along the $z$ axis, and $m$ signifies the mass of the ion.

Figure~\ref{ion trap}(b) shows our three-drive scheme: two second-sideband drives ($\omega_{r,b}=\omega_0\mp 2\omega +\delta_{r,b}$) and one carrier drive ($\omega_{\rm S} = \omega_0$ ), with phases $\phi_{{r,b}}=\frac{\pi}{2}$ and $\phi_{{\rm S}}=\pi$. Using the $4^{2}S_{1/2} \leftrightarrow 3^{2}D_{5/2}$ transition in $^{40}\text{Ca}^{+}$, we operate in the Lamb-Dicke regime ($\eta \sqrt{\langle n\rangle} \ll 1$, where $\langle n\rangle$ is the average phonon number) with phonons prepared to ground state via laser-cooling techniques with a high fidelity. The rotating-wave approximation yields the Hamiltonian:
\begin{eqnarray}\label{HLD}
\hat{H}_{\rm LD}&=&-i\frac{\eta^{2}\Omega_{r}}{2} \hat{a}^{2} \hat{\sigma}^+ e^{-i\delta_rt}  -i\frac{\eta^{2}\Omega_{b}}{2}  \hat{a}^{\dagger2} \hat{\sigma}^+ e^{-i\delta_bt}  \nonumber  \\
&&  -\frac{\Omega_{\rm S}}{2}\left(1-\frac{\eta^{2}}{2}-\eta^2\hat{a}^\dag \hat{a}\right)\hat{\sigma}^++{\rm H.c.},
\end{eqnarray}
where we consider the second order in $\eta$ and $\Omega_{0}=\Omega_{\rm S}\left(1-\frac{\eta^{2}}{2}\right) $.

To enhance the Stark coupling relative to the free energy of qubit and phonon, we implement unitary transformations: first move to an interaction picture with respect to $\frac{\Omega}{2}\hat{\sigma}_x$ (where $\Omega=-(\Omega_{0}+\omega_{q})$ and $\omega_{q}$ represents target qubit spacing), then apply a rotating-frame transformation via $-\omega_{c}\hat{a}^\dagger \hat{a}$, while setting the detunings as $\delta_{r,b}=\Omega\pm\omega_{c}$. This yields the effective Hamiltonian $\hat{H}^{\prime}_{\rm LD}=\omega_{c} \hat{a}^\dagger \hat{a}+ \frac{\omega_{q}}{2}\hat{\sigma}_x+ \lambda \sigma_y\left(\hat{a}^{2}+\hat{a}^{\dagger2}\right)+U \hat{a}^\dagger \hat{a} \hat{\sigma}_x$, where higher-order corrections(scaling as $\eta^{2}$ and oscillating at $\Omega$ or $\omega$) are negligible. A qubit basis rotation converts this to the standard two-photon Rabi-Stark model
\begin{equation}\label{tprs}
\hat{H}^{\prime \prime}_{\rm LD}=\omega_{c} \hat{a}^\dagger \hat{a}+ \frac{\omega_{q}}{2}\hat{\sigma}_z+ \lambda  \hat{\sigma}_x\left(\hat{a}^{2}+\hat{a}^{\dagger2}\right)+U  \hat{a}^\dagger \hat{a}\hat{\sigma}_z
\end{equation}
with $\lambda=(\eta^{2}\Omega_r/8)(1-2\epsilon_{\rm S})$ and $U=\eta^2\Omega_{\rm S}/2$, where $\epsilon_{\rm S}=\Omega_{\rm S}/\omega$ and $\Omega_b=\Omega_r(1-2\epsilon_{\rm S})/(1+2\epsilon_{\rm S})$. The derivation of Eq.~(\ref{tprs}) from Eq.~(\ref{HLD}) is provided in the Supplemental Material.

By generalizing this approach, the $N$-qubit two-photon Dicke-Stark model can be realized using a chain of $N$ trapped ions, as illustrated in Fig.~\ref{ion trap}(a). In this setup, the bosonic mode is represented by the collective motional mode of the ion chain. With identical qubit frequencies ($\omega_{q}^{j}=\omega_{q}$) and uniform couplings ($g_j=g$) achieved via either: (i) global longitudinal illumination (simpler but limited for large $N$), or (ii) individual transverse addressing (precise but complex), the two-photon Dicke-Stark Hamiltonian is:
\begin{eqnarray}\label{DiStSame}
\hat{\mathcal{H}}&=&\omega_{c} \hat{a}^\dagger \hat{a} +\frac{\omega_q}{2} \sum_{j=1}^N  \hat{\sigma}_z^j + \frac{g}{N}\sum_{j=1}^N  \hat{\sigma}_x^j \left( \hat{a}^2 + {\hat{a}^{\dagger2}} \right)  \nonumber  \\
&& +U \sum_{j=1}^N   \hat{a}^\dagger \hat{a}\hat{\sigma}_z^j.
\end{eqnarray}
The system consists of $N$ two-level ions $j = (0,1,2,. . .,N)$ with uniform coupling strength $g=\lambda N$. Notably, all parameters are independently tunable, with $g$ controlled by the red-sideband Rabi frequency $\Omega_{r}$ and $U$ by the carrier-drive Rabi frequency $\Omega_{\rm S}$. In sight of these features demonstrated above, optical switching between the NP to the SP can be accomplished by adjusting the pump field intensity. Figure~\ref{ion trap} demonstrates this optical switching through phonon number (c) and pseudospin (d), where both quantities change continuously for $U>0.0168$, confirming a second-order transition.

We use the ion chain's center-of-mass mode as our bosonic field. To realize the two-photon interaction in Eq.~\eqref{DiStSame} without exciting  unintended motional modes (particularly the nearest breathing mode at $\omega_2 = \sqrt{3}\omega$), our drives ($\omega_{r,b}=\omega_0\mp 2\omega +\delta_{r,b}$) require detunings $\delta_{r/b}$ that avoid $\omega_2$ and $2\omega_2$ (the breathing mode's first and second sidebands). This is achieved by choosing sufficiently large $\omega$. In addition, the population of the state $|g\rangle$ can be measured through electron shelving technique.

The experimental parameters for Fig.~\ref{order parameter} use a trapping frequency $\omega=(2\pi)\times4.98$ MHz, $\eta=0.1$, and carrier drive $\Omega_{\rm S}=(2\pi)\times120$ kHz, yielding Stark coupling $U=(2\pi)\times0.6$ kHz. For panel (b), the ratios  $U/\omega_{c}=0.0168$ and $g/\omega_{c}=0.33$ are achieved with $\Omega_r=(2\pi)\times200$ kHz, $\Omega_b=(2\pi)\times180$ kHz, and $\Omega=(2\pi)\times(-119.9)$ kHz, while panel (c) parameters $U/\omega_{c}=0.03$ and $g/\omega_{c}=0.22$ require $\Omega_r=(2\pi)\times70$ kHz, $\Omega_b=(2\pi)\times66$ kHz, and $\Omega=(2\pi)\times(-119.7)$ kHz. All of parameters are experimentally feasible.

\noindent{\it Conclusions. }
In summary, we have investigated quantum phase transitions in the two-photon Dicke-Stark model, which includes both Rabi and Stark couplings. We find that this system undergoes a second-order SPT. Importantly, the Stark coupling reduces the required critical Rabi coupling strength for transition, thereby making the SPT accessible even in strong coupling regimes.
Using the Holstein-Primakoff transformation combined with mean-field approximation, we derived analytical expressions for the ground-state energy and obtained the critical condition for this phase transition. The resulting atomic averages confirm that the SPT thresholds $g_{t}$ can be tuned by modulating the Stark coupling.
Beyond the mean-field theory, our analysis including quantum fluctuations yields low-energy effective Hamiltonians for both NP and SP, providing spin-squeezing fluctuation characteristics and further confirming the role of Stark coupling in tuning the SPT point. Notably, our approaches account for and avoids the spectral collapse typically associated with the two-photon Dicke model.
Furthermore, we propose a trapped-ion implementation of the two-photon Dicke-Stark model using a three-drive scheme comprising two second-sideband drives and one carrier drive. All the parameters in this scheme are experimentally feasible. As a practical application of this ion system, optical switching between the NP and the SP can be realized by adjusting the laser field intensity.
The results presented in this work demonstrate the potential for optical control of phase transitions. Moreover, our proposed scheme can be extended to various quantum systems, including the atoms and quantum dots cavity QED systems as well as the circuit QED system.

\acknowledgments
The authors acknowledge fruitful discussions with Le-Man Kuang, Xin-You L\"{u}, Jun Li, Jin-Feng Huang, Qing-Shou Tan, Ting Chen, Yi Xie and Jie Zhang. This work was supported by the National Science Foundation of China (Grants No. 12174448, No 12174447), and the Innovation Program for Quantum Science and Technology (Grant No. 2021ZD0301605).


\clearpage

\onecolumngrid
\vspace{\columnsep}
\begin{center}
\textbf{\large Supplemental Material}
\end{center}
\vspace{\columnsep}
\twocolumngrid

\setcounter{equation}{0}
\setcounter{figure}{0}
\setcounter{table}{0}
\setcounter{page}{1}
\makeatletter
\renewcommand{\theequation}{S\arabic{equation}}
\renewcommand{\thefigure}{S\arabic{figure}}
\renewcommand{\bibnumfmt}[1]{[S#1]}
\renewcommand{\citenumfont}[1]{S#1}

\section{formula for the order parameter}
To determine the ground-state value of $\beta$ (the order parameter), we minimize the system energy. In the thermodynamic limit, $\beta$'s behavior change signals a phase transition. For $g<g_t$, $E_G$ minimizes at $\beta=0$, while $g>g_t$ yields two degenerate minima at
\begin{eqnarray}
\beta &=& \sqrt{\frac{u_{1}[u_{0} + 64g^{2}(u_{4} -u_{3}) +16\omega^{2}(u_{3}-u_{5})]}{(16g^{2}+u_{2})^{2}(4g^{2}+u_{5}-u_{3})}}\nonumber  \\
&& +\frac{N(16g^{2} +u_{2} -2\omega_{c}\sqrt{u_{2}})}{2(16g^{2}+u_{2})},
\end{eqnarray}
and its negative number. For ease of notation, we define the following parameters:
\begin{subequations}\label{Ui}
\begin{equation}
u_{0}=u_{2}(4u_{5} -u_{2} -4u_{3} +4 \omega_{c}^{2} -16g^{2}),
\end{equation}
\begin{equation}
u_{1}=g^{2} N^{2},\hspace{5pt} u_{2}=U^{2} N^{2},\hspace{5pt} u_{3}=\omega_{q}^{2} N^{2},
\end{equation}
\begin{equation}
u_{4}=g \omega_{q} N^{2},\hspace{5pt} u_{5}=U \omega_{q} N^{2}.
\end{equation}
\end{subequations}

\section{Low-energy effective Hamiltonians}
\subsection{\label{sec:level2}normal phase}
In the first phase, $\beta=0$ indicates that we can substitute $\dop$ for $\bop$ in the calculation. With a reformulation, the Hamitonian of Eq.~(\ref{model DSM J}) neglecting terms of the order of $O(N^{-3/2}) $ can be simply expressed as
\begin{eqnarray}
\hat{H}_1&=&\Ko + \frac{\omega_1}{N}\opddag \dop + \frac{\omega_2}{\sqrt{N}}(\dop+\opddag)(\Kp + \Km)\nonumber \\
&&+\frac{\omega_{3}}{N} \opddag\dop\Ko,
\label{eqH1}
\end{eqnarray}
with the following definition
\begin{subequations}\label{omegan}
\begin{equation}
\omega_1=\frac{2\omega_q N-UN}{4\omega_{c}-2UN}, \hspace{5pt} \omega_2=\frac{2g}{2\omega_{c}-UN},
\end{equation}
\begin{equation}
\omega_3=\frac{2UN}{2\omega_{c}-UN}.
\end{equation}
\end{subequations}
Under the assumption that $U\sim\omega_{q}$ and accounting for the thermodynamic limit for atoms, $N\rightarrow \infty$, we get $\omega_{1}=O(1)$, $\omega_{2}=O(1)$, and $\omega_{3}=O(1)$. We consider a unitary transformation $\hat{U}=e^{-\hat{S}}$ where the generator
\begin{equation}
\hat{S}=-\frac{\omega_2}{\sqrt{N}}(\dop+\opddag)(\Kp-\Km).
\end{equation}

Using the Schrieffer-Wolff transformation, we expand the Hamiltonian in powers of $N^{-1/2}$ about the bosonic fluctuation operator $\dop$. This expansion yields
\begin{eqnarray}\label{eqH1prime}
\hat{H}_1\sp{\prime}&=&e^{-\Sop} \Hop_1 e^{\Sop} =\Ko + \frac{\omega_1}{N}\opddag \dop - \frac{2\omega_2^2}{N}(\dop+\opddag)^2\Ko \nonumber \\
&& + \frac{\omega_{3}}{N}\opddag \dop \Ko + O\left(\frac{1}{N\sqrt{N}}\right).
\end{eqnarray}
After projecting the Hamiltonian into the lowest-energy eigenspace of $\Ko$  with $\langle \Ko \rangle= 1/4$, the Hamiltonian of Eq.~(\ref{eqH1prime}) can be simply diagonalized by applying a squeezing operator $r_{s}^{(1)}=\ln(1-\frac{8g^{2}}{(2\omega_{c}-UN)\omega_{q}N})$.

\subsection{\label{sec:level2} superradiant phase}
In the second phase, $\beta $ is no longer equal to 0, and $\bop=\beta + \dop$. Applying the Holstein-Primakoff transformation and keeping terms up to $O(N^{-3/2})$, we obtain the simplified form of the Hamiltonian from Eq.~(\ref{model DSM J}):
\begin{eqnarray}
\Hop_{2} & = &  \omega_{q} \opddag \dop + \omega_{q} \beta (\opddag+\dop) + 2U \opddag \dop \Ko + 2U \beta (\opddag + \dop) \Ko \nonumber  \\
& & + \frac{4g \chi \delta}{\sqrt{N}}(\opddag + \dop) \hat{K}_{\text{X}} - \frac{2g \alpha}{N \chi}(\hat{d}^{\dag 2} +\hat{d}^{ 2}  + 4 \opddag \dop )\hat{K}_{\text{X}} \nonumber \\
 & & - \frac{g\alpha^{3}}{2N \chi^{3}}(\opddag + \dop)^{2} \hat{K}_{\text{X}} +\hat{H}_{f} + \hat{H}_{\text{Cons}},
\end{eqnarray}
where $\hat{H}_{\text{Cons}}=\omega_q\beta^{2} - \omega_qN/2- \omega_{c}/2$, $\hat{H}_{f}= 2\omega_{c}^{\prime} \Ko + (8g\beta\chi / \sqrt{N})\hat{K}_{\text{X}}$ and $\omega_{c}^{\prime}=\omega_{c} + U\beta^{2} -UN/2$. For ease of notation, we define the following parameters:
\begin{subequations}
\begin{equation}
\alpha=\frac{\beta}{\sqrt{N}}=O(1),
\end{equation}
\begin{equation}
\chi=\sqrt{1-\frac{\beta^2}{N}}=\sqrt{1-\alpha^2}=O(1),
\end{equation}
\begin{equation}
\delta=1-\frac{\beta^2}{N-\beta^2}=O(1).
\end{equation}
\end{subequations}
By removing the constant term and dividing it by a factor, we simplify the formula: $\hat{H}_{2}=(\hat{H} + \omega_qN/2 + \omega_{c}/2 -\omega_q\beta^{2})/2\omega_{c}^{\prime}$ and obtain:
\begin{eqnarray}
\Hop_{2} & = & \Ko +\frac{4g\alpha \chi}{\omega_{c}^{\prime}} \hat{K}_{\text{X}} + \frac{\omega_{q}}{2\omega_{c}^{\prime}} \opddag \dop + \frac{\omega_{q}\beta}{2\omega_{c}^{\prime}}  (\opddag+\dop)   \nonumber  \\
& & + \frac{U}{\omega_{c}^{\prime}} \opddag \dop \Ko + \frac{U \beta}{\omega_{c}^{\prime}}(\opddag + \dop) \Ko + \frac{2g \chi \delta}{\omega_{c}^{\prime}\sqrt{N}}(\opddag + \dop) \hat{K}_{\text{X}} \nonumber  \\
& &  - \frac{g \alpha}{\omega_{c}^{\prime} N \chi}(\hat{d}^{\dag 2} +\hat{d}^{ 2}  + 4 \opddag \dop )\hat{K}_{\text{X}} - \frac{g\alpha^{3}}{4\omega_{c}^{\prime}N \chi^{3}}(\opddag + \dop)^{2} \hat{K}_{\text{X}}. \nonumber \\
\end{eqnarray}
To diagonalize the field part of the above Hamiltonian $\Ko + (4g\alpha \chi/\omega_{c}^{\prime}) \hat{K}_{\text{X}}$, we apply a squeezing operator
\begin{equation}
r_a^{(2)} =  \frac{1}{2}\arctanh\left(\frac{g\alpha}{\omega_{c}^{\prime} N\chi} + \frac{4g\alpha \chi}{\omega_{c}^{\prime}}\right).
\end{equation}
Then the transformed Hamiltonian is derived as
\begin{eqnarray}
\Hop_{2} & = & \lambda_{0} \Kop + \frac{\lambda_{1}}{\sqrt{N}}(\dop+\opddag) + \frac{\lambda_{2}}{\sqrt{N}}(\dop+\opddag)(\Kpp + \Kmp)  \nonumber  \\
& & + \frac{\lambda_{3}}{\sqrt{N}}(\dop+\opddag)\Kop + \frac{\lambda_{4}}{N}\opddag \dop + \frac{\lambda_{5}}{N} \opddag \dop \Kop   \nonumber \\
 & & - \frac{2}{N}\Kop \hat{V}_{1}(\dop) + \frac{1}{N}(\Kpp + \Kmp) \hat{V}_{2}(\dop) \nonumber \\
& & - \frac{\lambda_{6}}{N}    \opddag \dop (\Kpp + \Kmp) ,
\end{eqnarray}
where we have defined new operators and parameters:
\begin{subequations}
\begin{equation}
\Kop =\cosh(x)\Ko + \frac{1}{2}\sinh(x)(\Kp + \Km),
\end{equation}
\begin{equation}
\Kpp + \Kmp = \cosh(x)(\Kp + \Km)+ 2\sinh(x)\Ko,
\end{equation}
\end{subequations}

\begin{subequations}
\begin{equation}
\lambda_0=\cosh(x) - \left(\frac{4g\alpha\chi}{\omega_{c}^{\prime}} + \frac{g\alpha}{\omega_{c}^{\prime}\chi N}\right)\sinh(x),
\end{equation}
\begin{equation}
\lambda_1=\frac{\omega_qN\alpha}{2\omega_{c}^{\prime}},
\end{equation}
\begin{equation}
\lambda_2 = \frac{g\chi\delta}{\omega_{c}^{\prime}}\cosh(x)-\frac{\alpha UN}{2\omega_{c}^{\prime}} \sinh(x),
\end{equation}
\begin{equation}
\lambda_3=-\frac{g\chi\delta}{2\omega_{c}^{\prime}}\sinh(x)+\frac{\alpha UN}{\omega_{c}^{\prime}} \cosh(x),
\end{equation}
\begin{equation}
\lambda_4=\frac{\omega_qN}{2\omega_{c}^{\prime}},
\end{equation}
\begin{equation}
\lambda_5=\frac{UN}{\omega_{c}^{\prime}} \cosh(x),
\end{equation}
\begin{equation}
\lambda_6=\frac{UN}{2\omega_{c}^{\prime}} \sinh(x),
\end{equation}
\begin{equation}
\hat{V}_1(\dop)= \sinh(x) \left(-\frac{g\alpha}{\chi \omega_{c}^{\prime}}\opddag \dop - \frac{g\alpha^{\prime}}{\omega_{c}^{\prime}} (\dop+\opddag)^2\right),
\end{equation}
\begin{equation}
\hat{V}_2(\dop)= \cosh(x) \left[-\left(\frac{g\alpha}{\chi \omega_{c}^{\prime}}+\lambda_6\right)\opddag \dop - \frac{g\alpha^{\prime}}{\omega_{c}^{\prime}}(\dop+\opddag)^2\right].
\end{equation}
\end{subequations}
Here, $\alpha^{\prime}=\alpha/(2\chi)+ \alpha^3/(4\chi^3)$ and $x=2r_a^{(2)}$. We seek to decouple the eigenspaces of $\Kop $; for this, we apply a transformation $e^{-\Sop} \Hop_2 e^{\Sop}$ with $\hat{S}=\frac{1}{\sqrt{N}}\Sop_1 +\frac{1}{N} \Sop_2$. The operators are introduced as:
\begin{subequations}

\begin{equation}
\Sop_1=-\frac{\lambda_2}{\lambda_0}(\dop+\opddag)(\Kpp - \Kmp),
\end{equation}
\begin{equation}
\Sop_2=(\Kpp - \Kmp) \left(\frac{\lambda_3\lambda_2}{\lambda_0^2}(\dop+\opddag)^2 - \frac{\hat{V}_2(\dop)}{\lambda_0}\right).
\end{equation}
\end{subequations}
This gives us a Hamiltonian that commutes with $\Kop$:
\begin{eqnarray}
\Hop_{2}^{\prime} & = & e^{-\Sop} \Hop_2 e^{\Sop} \nonumber  \\
&=& \lambda_0\Kop  + \frac{\lambda_4}{N}\opddag \dop + \frac{\lambda_1}{\sqrt{N}}(\dop+\opddag)  \nonumber  \\
& & - \frac{2\lambda_2^2}{N\lambda_0}(\dop+\opddag)^2\Kop + \frac{\lambda_3}{\sqrt{N}}(\dop+\opddag)\Kop  \nonumber  \\
& &  - \frac{2}{N}\Kop \hat{V}_1(\dop) .
\end{eqnarray}
Projection in the ground state of $\Kop$ gives $\langle \Kop \rangle= 1/4$. By selecting an appropriate $\alpha$, the first-order nonlinear term can be eliminated. Then restoring the constants in the Hamiltonian yields the  Eq.~(\ref{Hsp}) described in the main text:
\begin{equation}
\Hop_{SP}  =  \frac{\lambda_0}{4}  +\omega_{1}\opddag \dop +\omega_{2}(\dop+\opddag)^2.
\end{equation}
It can be diagonalized by a Bogoliubov transformation of the squeezing parameter
\begin{equation}
r_{s}^{(2)}=-\frac{1}{4}\ln(1+\frac{\omega_{2}}{\omega_{1}}),
\end{equation}
where
\begin{subequations}
\begin{equation}
\omega_{1}=\frac{\lambda_4}{N} +\frac{g\alpha}{2\omega_{c}^{\prime}\chi N}\sinh(x) +\frac{\lambda_5}{4N} ,
\end{equation}
\begin{equation}
\omega_{2}=\frac{g \alpha^{\prime}}{2\omega_{c}^{\prime} N} \sinh(x)- \frac{\lambda_2^2}{2N\lambda_0}.
\end{equation}
\end{subequations}

\begin{widetext}

\section{Derivation of the Rabi-Stark Hamiltonian in trapped ions}
In this section, we will explain in detail how to derive the effective Hamiltonian in Eq.~(\ref{tprs}) from Eq.~(\ref{HLD}),
\begin{equation}\label{SScheme2}
 \hat{H}_{A}(t)=-i\frac{\eta^{2}\Omega_r}{4} \hat{a}^{2} \hat{\sigma}_+ e^{-i\delta_rt}  -i\frac{\eta^{2}\Omega_b}{4}  {\hat{a}^{\dagger2}} \hat{\sigma}_+ e^{-i\delta_bt} - g_{\rm S} \hat{\sigma}_++{\rm H.c.}.
\end{equation}
Here, $g_{\rm S}=\frac{\Omega_{\rm S}}{2}(1-\eta^2/2)-\frac{\Omega_{\rm S}}{2}\eta^2\hat{a}^\dag \hat{a}=\frac{\Omega_0}{2}-U \hat{a}^\dagger \hat{a}$. For a complete description of the effective dynamics beyond the rotating-wave approximation (RWA), the total Hamiltonian must include additional terms: $ \hat{H}= \hat{H}_A+ \hat{H}_B$, where
\begin{equation}\label{SSchemeRemaining}
 \hat{H}_{B}(t)=-\frac{\eta \Omega_r}{2} \hat{\sigma}_+ \hat{a} e^{-i(-\omega +\delta_r)t} -\frac{\eta \Omega_b}{2} \hat{\sigma}_+ \hat{a}^{\dagger} e^{-i(\omega +\delta_b)t} - i\frac{\eta \Omega_{\rm S}}{2}\hat{\sigma}_+(\hat{a}e^{-i\omega t}+\hat{a}^\dagger e^{i\omega t}) +{\rm H.c.}.
\end{equation}
The first two terms are the off-resonant interactions of the red and blue drivings which are usually ignored given that $\Omega_{r,b}\ll \omega$. The third term represents the coupling from the carrier driving. As we will see in the following, these non-commuting terms rotate at comparable frequencies $\delta_r,\delta_b\ll\omega$, generating significant second-order interactions. The resulting second-order effective Hamiltonian is:
\begin{equation}\label{SSecondOrder}
 \hat{H}^{(2)}(t)=-i\int_{0}^{t} \hat{H}(t) \hat{H}(t')dt'=-i\int_{0}^{t}\Big( \hat{H}_A(t)+ \hat{H}_B(t)\Big)\Big( \hat{H}_A(t')+ \hat{H}_B(t')\Big)dt'.
\end{equation}

We are only interested in terms arising from $\int_{0}^{t} \hat{H}_B(t) \hat{H}_B(t')dt'$ whose oscillating frequency is $\delta_r$ or $\delta_b$. These are
\begin{eqnarray}\label{SSecondOrderList}
-\frac{\eta\Omega_r}{2}\hat{\sigma}_+ \hat{a} e^{-i(-\omega+\delta_r)t}\int_{0}^tdt'(i\eta)\frac{\Omega_{\rm S}}{2}\hat{\sigma}_- \hat{a}e^{-i\omega t'} =\eta^{2}\frac{\Omega_{\rm S}\Omega_r}{4\omega}\hat{\sigma}_+\hat{\sigma}_-e^{-i\delta_r t} \hat{a}^{2}, \\
-\frac{\eta\Omega_b}{2}\hat{\sigma}_+ \hat{a}^{\dagger} e^{-i(\omega+\delta_b)t}\int_{0}^tdt'(i\eta)\frac{\Omega_{\rm S}}{2}\hat{\sigma}_- \hat{a}^\dagger e^{i\omega t'} =-\eta^{2}\frac{\Omega_{\rm S}\Omega_b}{4\omega}\hat{\sigma}_+\hat{\sigma}_-e^{-i\delta_b t} \hat{a}^{\dagger2}, \\
-i\eta\frac{\Omega_{\rm S}}{2}\hat{\sigma}_+ \hat{a}^\dagger e^{i\omega t}\int_0^t dt' (-\frac{ \eta\Omega_r}{2})\hat{\sigma}_- \hat{a}^\dagger e^{i(-\omega+\delta_r)t'}=-\eta^{2}\frac{\Omega_{\rm S}\Omega_r}{4(\omega-\delta_r)}\hat{\sigma}_+\hat{\sigma}_-e^{i\delta_r t} \hat{a}^{\dagger2}, \\
-i\eta\frac{\Omega_{\rm S}}{2}\hat{\sigma}_+ \hat{a} e^{-i\omega t}\int_0^t dt' (-\frac{\eta\Omega_b}{2})\hat{\sigma}_- \hat{a} e^{i(\omega+\delta_b)t'}=\eta^{2}\frac{\Omega_{\rm S}\Omega_b}{4(\omega+\delta_b)}\hat{\sigma}_+\hat{\sigma}_-e^{i\delta_b t} \hat{a}^{2}, \\
-\frac{\eta\Omega_r}{2}\hat{\sigma}_- \hat{a}^{\dagger} e^{i(-\omega+\delta_r)t}\int_{0}^tdt'(-i\eta)\frac{\Omega_{\rm S}}{2}\hat{\sigma}_+ \hat{a}^\dagger e^{i\omega t'} =\eta^{2}\frac{\Omega_{\rm S}\Omega_r}{4\omega}\hat{\sigma}_-\hat{\sigma}_+e^{i\delta_r t} \hat{a}^{\dagger2}, \\
-\frac{\eta\Omega_b}{2}\hat{\sigma}_- \hat{a} e^{i(\omega+\delta_b)t}\int_{0}^tdt'(-i\eta)\frac{\Omega_{\rm S}}{2}\hat{\sigma}_+ \hat{a} e^{-i\omega t'} =-\eta^{2}\frac{\Omega_{\rm S}\Omega_b}{4\omega}\hat{\sigma}_-\hat{\sigma}_+e^{i\delta_b t} \hat{a}^{2}, \\
i\eta\frac{\Omega_{\rm S}}{2}\hat{\sigma}_- \hat{a} e^{-i\omega t}\int_0^t dt' (-\frac{\eta\Omega_r}{2})\hat{\sigma}_+ \hat{a}e^{-i(-\omega+\delta_r)t'}=-\eta^{2}\frac{\Omega_{\rm S}\Omega_r}{4(\omega-\delta_r)}\hat{\sigma}_-\hat{\sigma}_+e^{-i\delta_r t} \hat{a}^{2},  \\
i\eta\frac{\Omega_{\rm S}}{2}\hat{\sigma}_- \hat{a}^\dagger e^{i\omega t}\int_0^t dt' (-\frac{\eta\Omega_b}{2})\hat{\sigma}_+ \hat{a}^{\dagger}e^{-i(\omega+\delta_b)t'}=\eta^{2}\frac{\Omega_{\rm S}\Omega_b}{4(\omega+\delta_b)}\hat{\sigma}_-\hat{\sigma}_+e^{-i\delta_b t} \hat{a}^{\dagger2}.
\end{eqnarray}
If we assume that $1/(\omega\pm\delta_j)\sim1/\omega $ and reorganize all the terms, we get that the second-order effective Hamiltonian is
\begin{equation}\label{SSecondOrderEff}
\hat{H}_B^{(2)}(t)\approx -i\eta^{2}\frac{\Omega_{\rm S}\Omega_r}{4\omega} (e^{-i\delta_r t} \hat{a}^{2}+{\rm H.c.})\hat{\sigma}_z + i\eta^{2}\frac{\Omega_{\rm S}\Omega_b}{4\omega} (e^{-i\delta_b t} \hat{a}^{\dagger2}+{\rm H.c.})\hat{\sigma}_z,
\end{equation}
which can be incorporated to the first-order Hamiltonian in Eq.~(\ref{SScheme2}), giving
\begin{equation}\label{SEffectiveQRS}
\hat{H}_{\rm eff}(t)=-(2g^{(1)}_r \hat{\sigma}_+ +g^{(2)}_{r}\hat{\sigma}_z) \hat{a}^{2}e^{-i\delta_rt}  -(2g_b^{(1)}\hat{\sigma}_+ -g_{b}^{(2)}\hat{\sigma}_z) \hat{a}^{\dagger2} e^{-i\delta_bt} - \frac{\Omega_0}{2} \hat{\sigma}_+  +\eta^2\frac{\Omega_{\rm S}}{2} \hat{a}^\dagger \hat{a}\hat{\sigma}_+ +{\rm H.c.},
\end{equation}
where $g_{r,b}^{(1)}=i\eta^{2}\Omega_{r,b}/8$, $g_{r,b}^{(2)}=i\eta^{2}\Omega_{\rm S}\Omega_{r,b}/4\omega$ and $\Omega_0=\Omega_{\rm S}(1-\eta^2/2)$. Now, move to a frame w.r.t. $\hat{U}=\text{exp}(-i\frac{\Omega}{2}\hat{\sigma}_{x})$, i.e. $\hat{H}_{\rm eff}^{I}= \hat{U} \hat{H}_{\rm eff}(t) \hat{U}^{\dagger}$. Then, we obtain
\begin{equation}\label{SEffectiveQRS2}
\hat{H}_{\rm eff}^{I}= \frac{\omega_{q}}{2}\hat{\sigma}_+ + \eta^2\frac{\Omega_{\rm S}}{2} \hat{a}^\dagger \hat{a}\hat{\sigma}_+ -\Big(2g^{(1)}_r \hat{U}^{\dagger}\hat{\sigma}_+ \hat{U} + g^{(2)}_{r} \hat{U}^{\dagger}\hat{\sigma}_z \hat{U}\Big) \hat{a}^{2}e^{-i\delta_rt} -\Big(2g_b^{(1)}\hat{U}^{\dagger}\hat{\sigma}_+ \hat{U} - g_{b}^{(2)}\hat{U}^{\dagger}\hat{\sigma}_z \hat{U} \Big) \hat{a}^{\dagger2} e^{-i\delta_bt} +{\rm H.c.},
\end{equation}
where $\omega_{q}=-\Omega_0-\Omega$. Using that
\begin{eqnarray}\label{SRotatingPaulis}
\hat{U} \hat{\sigma}_y \hat{U}^{\dagger}&=&\tilde{\sigma}_+e^{ i\Omega t} + \tilde{\sigma}_-e^{- i\Omega t},\\
\hat{U} \hat{\sigma}_z \hat{U}^{\dagger}&=& -i(\tilde{\sigma}_+e^{ i\Omega t} - \tilde{\sigma}_-e^{- i\Omega t}),
\end{eqnarray}
where $\tilde{\sigma}_{\pm}=(\hat{\sigma}_y\pm i\hat{\sigma}_z)/2$, and that the detunings are chosen to be $\delta_r=\Omega +\omega_{c}$ and $\delta_b=\Omega -\omega_{c}$, Eq.~(\ref{SEffectiveQRS}) is rewritten as
\begin{eqnarray}\label{SEffectiveQRS3}
\hat{H}_{\rm eff}^{I}= \frac{\omega_{q}}{2}\hat{\sigma}_+ + \eta^2\frac{\Omega_{\rm S}}{2} \hat{a}^\dagger \hat{a}\hat{\sigma}_+ + \Big(g^{(1)}_r (-\hat{\sigma}_x - i\tilde{\sigma}_+ e^{i\Omega t} -i \tilde{\sigma}_-e^{- i\Omega t}) + g^{(2)}_{r}(i\tilde{\sigma}_+e^{i\Omega t} -i \tilde{\sigma}_-e^{- i\Omega t})\Big) \hat{a}^{2}e^{- i\Omega  t} e^{- i\omega_{c} t} \nonumber\\
 +\Big(g_b^{(1)}(-\hat{\sigma}_x -i\tilde{\sigma}_+e^{ i\Omega t} -i \tilde{\sigma}_-e^{- i\Omega t}) - g_{b}^{(2)}(i\tilde{\sigma}_+e^{ i\Omega t} - i \tilde{\sigma}_-e^{- i\Omega t})\Big) \hat{a}^{\dagger2} e^{- i\Omega  t}e^{ i\omega_{c} t} +{\rm H.c.},
\end{eqnarray}
where all terms rotating with $\pm\Omega$ or higher can be ignored using the RWA. After the approximation we have that
\begin{equation}\label{SEffectiveQRS4}
\hat{H}_{\rm eff}^{I}= \frac{\omega_{q}}{2}\hat{\sigma}_+ + \eta^2\frac{\Omega_{\rm S}}{2} \hat{a}^\dagger \hat{a}\hat{\sigma}_+ -i (g^{(1)}_r  - g^{(2)}_{r})\tilde{\sigma}_+ \hat{a}^{2} e^{- i\omega_{c} t}  -i(g_b^{(1)} + g_{b}^{(2)})\tilde{\sigma}_+  \hat{a}^{\dagger2} e^{ i\omega_{c} t} +{\rm H.c.},
\end{equation}
which, in a rotating frame w.r.t $-\omega_{c} \hat{a}^\dagger \hat{a}$, transforms to
\begin{equation}\label{SEffectiveQRS5}
\hat{H}_{\rm eff}^{II}= \frac{\omega_{q}}{2}\hat{\sigma}_x +\omega_{c} \hat{a}^\dagger \hat{a} + g_{\rm JC}(\tilde{\sigma}_+ \hat{a}+\tilde{\sigma}_- \hat{a}^\dagger)   +g_{\rm aJC}(\tilde{\sigma}_+  \hat{a}^\dagger+\tilde{\sigma}_- \hat{a}) \mp \eta^2\frac{\Omega_{\rm S}}{2} \hat{a}^\dagger \hat{a}\hat{\sigma}_x,
\end{equation}
where $g_{\rm JC}=\eta^{2}\Omega_r(1- 2\Omega_{\rm S}/\omega)/8$ and $g_{\rm aJC}=\eta^{2}\Omega_b(1+ 2\Omega_{\rm S}/\omega)/8$. With the parameter choice $\Omega_b=\Omega_r(1-2\epsilon_{\rm S})/(1+2\epsilon_{\rm S})$ (where $\epsilon_{\rm S}=\Omega_{\rm S}/\omega$) and a qubit basis transformation, we obtain the Eq.~(\ref{tprs}) in the main text.

\end{widetext}

\end{document}